\documentclass[prl,aps,twocolumn,superscriptaddress,showpacs]{revtex4}
\usepackage{amsbsy}
\usepackage{amssymb}
\usepackage[dvips]{graphicx}

\begin{document}
\title{Magnetic exponents of two-dimensional Ising spin glasses}

\author{F. Liers}
\affiliation{Institut f{\"u}r Informatik, Universit{\"a}t zu K{\"o}ln,
Pohligstra{\ss}e 1, D-50969 K{\"o}ln,Germany.}

\author{O. C. Martin}
\affiliation{Univ Paris-Sud, UMR8626, LPTMS, Orsay, F-91405; CNRS, Orsay, 
F-91405, France.}

\date{\today}

\begin{abstract}
The magnetic critical properties of two-dimensional 
Ising spin glasses are controversial.
Using exact ground state determination,
we extract the properties of clusters flipped
when increasing continuously a uniform field. We show that these clusters
have many holes but otherwise have statistical properties 
similar to those of zero-field droplets. A detailed analysis
gives for the magnetization exponent 
$\delta \approx 1.30 \pm 0.02$ using
lattice sizes up to $80 \times 80$; this is 
compatible with the droplet model prediction $\delta = 1.282$.
The reason for previous disagreements stems from the need to analyze both
singular and analytic contributions in the low-field regime.
\end{abstract}

\maketitle

Spin glasses~\cite{MezardParisi87b,Young98} have been the focus of 
much interest because of their many remarkable features:
they undergo a subtle freezing transition as temperature is lowered,
their relaxational dynamics is slow (non-Arrhenius), they 
exhibit ageing, memory effects, etc.
Although there are
still some heated disputes concerning three-dimensional
spin glasses, the case of two dimensions is relatively consensual,
at least in the absence of a magnetic field. Indeed,
two recent 
studies~\cite{KatzgraberLee04,HartmannHoudayer04}
found that the thermal properties of two-dimensional
Ising spin glasses with Gaussian couplings agreed very well
with the predictions of the scaling/droplet 
pictures~\cite{BrayMoore86,FisherHuse86}. Interestingly,
the situation in the presence of a magnetic field remains
unclear; in particular, some Monte Carlo 
simulations~\cite{KinzelBinder83}
and basically all ground state 
studies~\cite{KawashimaSuzuki92,Barahona94,RiegerSanten96} seem 
to go against the scaling/droplet pictures. Nevertheless, since 
spin glasses often have large corrections to scaling,
the apparent disagreement with the 
droplet picture resulting from these studies may be misleading and 
tests in one dimension give credence to this claim~\cite{CarterBray03}.

In this study we use state of the art algorithms for determining
exact ground states in the presence of a magnetic field and
treat significantly larger lattice sizes than in previous work.
By finding the precise points where the ground states change
as a function of the field, we extract the excitations
relevant in the presence of a field which can then be
compared to the zero-field droplets.
Although for small size lattices we agree
with previous studies, at our larger ones 
a careful analysis, taking into account both the analytic
and the singular terms, gives excellent agreement with the droplet
picture.

\paragraph*{The model and its properties ---} We work on an $L \times L$
square lattice having Ising spins on its sites and
couplings $J_{ij}$ on its bonds. The 
Hamiltonian is
\begin{equation}
\label{eq:H}
H(\{\sigma_{i}\}) \equiv
  -\sum_{\langle ij \rangle} J_{ij}\sigma_{i}\sigma_{j}
  - B\sum_{i} \sigma_{i}
\end{equation}
The first sum runs over all nearest neighbor sites using
periodic boundary conditions to minimize
finite size effects. 
The $J_{ij}$ are independent random variables of
either Gaussian or exponential distribution.

It is generally agreed that two-dimensional spin glasses
have a unique critical point at $T=B=0$. There, 
the free energy is non-analytic and in fact,
standard arguments~\cite{Cardy96} suggest that
as $T \to 0$ and $B \to 0$ the free energy goes as
$\beta F(L,\beta) \sim \beta E_0 + G_s(T L^{y_T},B L^{y_B})$
where $E_0$ is the ground-state energy, $\beta$ the inverse temperature, while
$y_T$ and $y_B$ are the thermal and magnetic exponents.
Previous work when $B=0$ is compatible with this
form and in fact also agrees with the scaling/droplet picture
of Ising spin glasses in which one has 
$y_T=-\theta\approx 0.282$.
The stumbling block concerns the behavior when $B\ne 0$;
there, the droplet prediction in general dimension $d$ is
\begin{equation}
\label{eq:y_B}
y_B = y_T + d/2
\end{equation}
but the numerical evidence for this is muddled
at best. It is thus worth 
reviewing the hypotheses assumed by the droplet model
so that they can be tested directly.

We begin with the fact that in any dimension $d$,
a magnetic field destabilizes
the ground state beyond a characteristic length scale $\xi_B$.
To see this, consider an infinitesimal field and 
zero-field droplets of scale $\ell$.
These are expected to be compact. 
The interfacial energy of such droplets is $O(\ell^{\theta})$
while their total magnetization goes as 
$\ell^{d/2}$. The magnetic and
interfacial energy are then balanced when
$B$ reaches a value $O(1/\ell^{d/2-\theta})$: at that value 
of the field, some of the droplets will flip and the ground
state will be destabilized. We then see that for each
field strength there is an associated magnetic length scale
$\xi_B$
\begin{equation}
\xi_B \approx B^{-\frac{1}{{\frac{d}{2}-\theta}}}
\end{equation}
This leads to the identification $y_B = d/2 - \theta$
in agreement with Eq.~\ref{eq:y_B}, giving 
$y_b \approx 1.282$ at $d=2$.

The droplet model also predicts the scaling
of the magnetization in the $B \to 0$ limit via
the exponent $\delta$:
\begin{equation}
m(B) \sim B^{1/\delta}
\end{equation}
If this form also holds for infinitesimal
fields at finite $L$,
we can consider the field $B^*$ for which 
system-size droplets flip; this happens when $B = O(1/L^{y_B})$
and then the magnetization is $O(L^{-d/2})$, the droplets
having random magnetizations. This leads to
$m(B^*) \sim L^{-d/2}$ and $m(B^*) \sim [1/L^{y_B} ]^{1/\delta}$ so that
\begin{equation}
d \delta = 2 y_B
\end{equation}

Although the droplet model arguments are not proofs,
they seem quite convincing. Nevertheless, the
numerical studies measuring $\delta$ do not give 
good agreement with the prediction $\delta =1.282$.
For instance, using Monte Carlo at ``low enough'' temperatures,
Kinzel and Binder~\cite{KinzelBinder83}
find $\delta \approx 1.39$. Since thermalization
is difficult at low temperatures, it is preferable to
work directly with ground states, at least when that 
is possible. This was done by three independent 
groups~\cite{KawashimaSuzuki92,Barahona94,RiegerSanten96}
with increasing power, leading
to $\delta \approx 1.48$, $\delta \approx 1.54$ 
and $\delta \approx 1.48$. Taken together, these
studies show a real discrepancy with the droplet
prediction. 
To save the droplet model from this thorny situation, one can appeal
to large corrections to scaling. 
Such potential effects have been considered~\cite{CarterBray03}
in dimension one where it was shown
that $\xi_B$ was poorly fitted by a pure power law
unless the fields were very small. Here we 
revisit the two-dimensional case
to reveal either the size of the corrections
to scaling or a cause for the break down of the
droplet reasoning.

\paragraph*{Computation of ground states ---}

We determine the exact ground state of the Hamiltonian (\ref{eq:H}) by
computing a maximum cut in the graph of
interactions~\cite{Barahona82}, a prominent problem in
combinatorial optimization. Whereas it is polynomially solvable on
two-dimensional grids without a field
and couplings bounded by a polynomial in the size of the input,
it is NP-hard with an 
external field. In practice, we rely on a branch-and-cut
algorithm~\cite{BarahonaGrotschel88,LiersJunger04}.

\begin{figure}[h]
\includegraphics[width=8.5cm, height=5cm,angle=0]{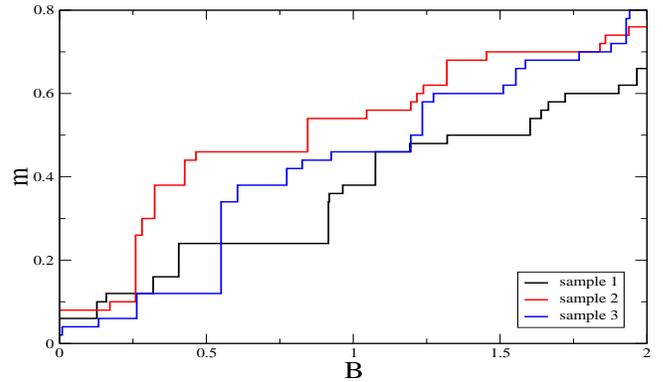}
\caption{Magnetization as a function of $B$ for three
typical $L=10$ samples.\protect\label{fig:fig1}}
\end{figure}

Let the ground state at a field $B$ be denoted as
$\{\sigma^{(G)}(B)\}$. To study the magnetization,
we computed the ground states at increasing values of $B$,
in steps of size $0.02$. When focusing instead on
the flipping clusters, we had to determine 
the intervals in which the ground state was constant
and in what manner it changed when going from one
interval to the next. In Fig.~\ref{fig:fig1} 
we show the associated piecewise constant magnetization curves
for three samples of the disorder variables $J_{ij}$ at $L=10$.
To get the sequence of intervals or break points
associated with such a function
exactly, we start by computing the ground state in zero field. By
applying postoptimality analysis from linear programming
theory, we determine~\cite{RiegerSanten96,LiersJunger04} a range
$\Delta B$ such that the ground state at a field $B$ remains the
optimum in the interval $[B,B+\Delta B]$. We reoptimize at 
$B+\Delta B+\epsilon$, with $\epsilon$ being a sufficiently small number. By
repeatedly applying this procedure, we get a new ground state
configuration, and increase $B$ until all spins are aligned with the
field. This procedure works for system sizes in which the
branch-and-cut program can prove optimality without branching, i.e.,
without dividing the problem into smaller sub-problems.
If the algorithm branches (this occurs
only for the largest system sizes studied here), we apply a
divide-and-conquer strategy for determining 
$\{\sigma^{(G)}(B)\}$ in an interval, say $[B_1,B_2]$. For a fixed
configuration the Hamiltonian (\ref{eq:H}) is linear in the field, the
slope being the system's magnetization. Let $f_1,f_2$ be the two
linear functions associated with $\{\sigma^{(G)}(B_1)\}$ and
$\{\sigma^{(G)}(B_2)\}$. If $f_1$ and $f_2$ are equal,
we are done. Otherwise, we determine the field $B_3$ at which the
functions intersect and recursively solve the problem in the intervals
$[B_1,B_3]$ and $[B_3,B_2]$. 

A typical sample at $L=80$ requires about 2 hours of cpu on a work
station for determining the ground states when $B$ goes through
the multiples of $0.02$. The more time consuming computation
of the exact break points takes about 4 hours 
on typical samples with $L=60$, but less than a minute if $L\le 30$
because the ground-state determinations are fast and branching almost
never arises. For our work, we considered mainly the
case of Gaussian $J_{ij}$, analyzing 2500 samples at $L=80$, 5000 at
$L=70$, and from 2000 to 11000 instances for sizes $L=60, 50, 40, 30,
24, 20, 14$. 
We also analyzed a smaller number of samples for $J_{ij}$
taken from an exponential distribution; exponents showed
no significant differences when comparing to the
Gaussian case.

\paragraph*{The exponent $\delta$ ---}
Given the Hamiltonian, it is easy to see that
for each sample the magnetization (density)
\begin{equation}
m_J(B) = \frac{\sum_i \sigma_i^{(G)}}{L^2} 
\end{equation}
must be an increasing function of $B$. (The index $J$ on 
the magnetization is to recall that it depends on the
disorder realisation, but in the large $L$ limit
$m_J$ is self averaging; also, without loss of generality,
we shall work with $B>0$.) At large fields $m_J$ saturates
to $1$, while at low fields, its growth law must be above 
a linear function of $B$. Indeed, for continuous $J_{ij}$, 
the distribution of local fields has a finite density at zero and so 
small clusters of spins will flip and will lead to a linear
contribution to the magnetization. A more singular
behavior is in fact predicted by the droplet model
since $\delta > 1$, indicating that 
the system is anomalously sensitive to the magnetic field perturbation.

\begin{figure}[h]
\includegraphics[width=8.5cm, height=6cm]{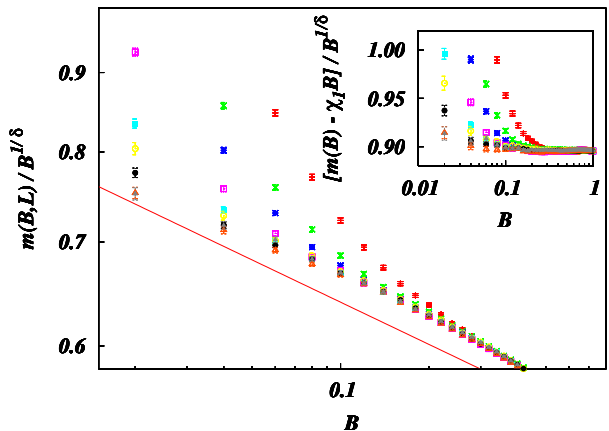}
\caption{Magnetization divided by $B^{1/\delta}$ as a function of $B$;
the $B^{1/1.45}$ line is to guide the eye. From top to bottom,
$L=14$, 20, 24, 30, 40, 50, 60, 70, and 80.
Inset: $m-\chi_1 B$ divided by 
$B^{1/\delta}$ as a function of $B$. (Same $L$ and
symbols as in main
part of the figure.) In both cases, $\delta$ is set to
its droplet model value, $\delta=1.282$.\protect\label{fig:fig2}}
\end{figure}

If $B$ is not too small, the convergence to the thermodynamic
limit ($L \to \infty$) is rapid, and in fact one expects 
exponential convergence in $L/\xi_B$. We should thus see
an envelope curve $m(B)$ appear as $L$ increases; to make
a power dependence on $B$ manifest, we show in Fig.~\ref{fig:fig2}
a log-log plot of the ratio $m(B)/B^{1/\delta}$ where 
$\delta$ is set to its droplet scaling value of 1.282.
For that value of $\delta$ there is not much indication that a flat
region is developping when $L$ increases, while at $L=50$ a direct fit
to a power gives $\delta = 1.45$ (cf. line displayed
in the figure to guide the eye), as found in previous
work~\cite{KawashimaSuzuki92,Barahona94,RiegerSanten96}.
The problem with this simple
analysis is that $m$ has both analytic and non-analytic contributions;
to lowest order we have
\begin{equation}
\label{eq:m_expansion}
m = \chi_1 B + \chi_S B^{1/\delta} + \ldots
\end{equation}
Although $\chi_1 B$ is sub-dominant, it is far from negligible
in practice; for instance for it to contribute to less than
$10\%$ of $m$, one would need $B < (0.1 \chi_S/\chi_1)^{1/0.282}$.
This could easily mean $B < 10^{-3}$ for which there would be huge finite size
effects since $L$ would then be much smaller than the magnetic length
$\xi_B$. We thus must take into account the term $\chi_1 B$; we have
done this, adjusting $\chi_1$ so that $(m - \chi_1 B)/B^{1/\delta}$
has an envelope as flat as possible. The result is displayed
in the inset of Fig.~\ref{fig:fig2}, showing that the 
droplet scaling fits very well the data as long as 
the $\chi_1 B$ term is included. In fact, direct fits to
the form of Eq.~\ref{eq:m_expansion}
give $\delta$ in the range 1.28 to 1.32 depending on the sets of $L$'s
included in the fits.

\paragraph*{The clusters that flip are like zero-field droplets ---}
The fundamental hypothesis
in the droplet argument relating $\delta$ or $y_B$ to $\theta$ is the fact
that in an infinitesimal field one flips droplets defined
in zero field, droplets which are compact and
have random (except for the sign) magnetizations. We 
therefore now focus on the
properties of the actual clusters that are flipped at low fields.

At zero field, the droplet of lowest energy 
almost always is a single spin (this
follows from the large number of such droplets, 
in spite of their typically higher energy).
Thus as the field is turned on, the ground state changes
first mainly via single spin flips, and
when large clusters do flip (they finally 
do so but at larger fields), 
they necessarily have many ``holes''
and thus do not correspond exactly to
zero-field droplets. This is not a problem for the droplet
argument as long as these clusters are
compact and have random magnetizations.
\begin{figure}[h]
\includegraphics[width=8.5cm, height=5cm]{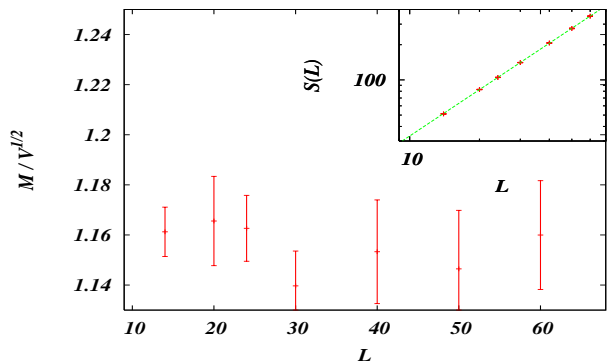}
\caption{The cluster magnetization divided
by the square root of cluster volume --- for the largest cluster flipped
in each sample --- is insensitive to $L$.
Inset: The clusters's mean surface scales as $L^{d_S}$ 
with $d_S\approx 1.32$.\protect\label{fig:fig3}}
\end{figure}
To test this, we consider for each realization of the $J_{ij}$ disorder
the \emph{largest} cluster that flips during
the whole passage from $B=0$ to $B=\infty$.
According to the droplet picture, this cluster
should contain a number of spins $V$ that scales
as $L^2$ (compactness) and have a total magnetization
$M$ that scales as $\sqrt{V}$ (randomness).
This is confirmed by our data where we find $M/V\sim 2/L$;
in Fig.~\ref{fig:fig3} we plot the disorder 
mean of $M/\sqrt{V}$ for increasing $L$; manifestly,
this mean is remarkably insensitive to $L$.
Similar conclusions apply to $V/L^2$. For completeness,
we show in the inset of the figure that the surface of 
these clusters, defined as the number of lattice
bonds connecting them to their complement,
grows as $L^{d_S}$ with $d_S\approx 1.32$;
this is to be compared to the value $d_S=1.27$ for
zero-field droplets~\cite{HartmannYoung02}, in spite
of the fact that our clusters have holes.
All in all, we find that the clusters considered have statistical
properties that are completely compatible with those assumed in
the droplet scaling argument, thereby directly validating
the associated hypotheses.

\begin{figure}[h]
\includegraphics[width=8.5cm, height=6cm]{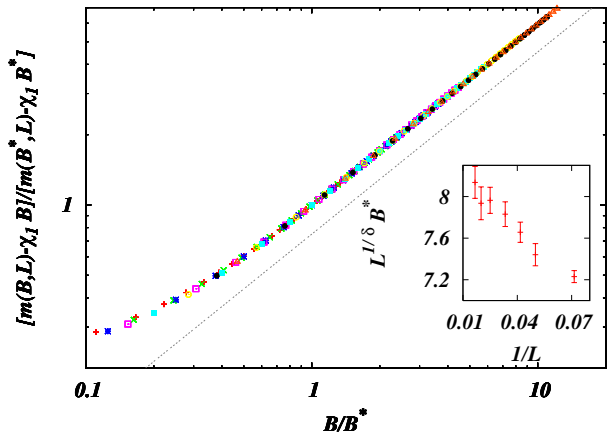}
\caption{Inset: Field $B^*$ times $L^{1/\delta}$
as a function of $1/L$ shows
a limit at large $L$
as expected in the droplet model ($\delta=1.282$).
Main figure: 
Data collapse plot exhibiting finite size scaling of the
singular part of the magnetization 
($L=10$, 14, 20, 24, 30, 40, 50, and 60).
\protect\label{fig:fig4}}
\end{figure}

\paragraph*{The magnetic exponent $y_B$  and finite size scaling of the magnetization ---}
One can also measure the exponent $y_B$ directly
via the magnetic length which scales as  $\xi_B \sim B^{-1/y_B}$.
For each sample, define $B_J^*$ as that field where the ground
state changes by the largest cluster of spins as described in the
previous paragraph. Since these clusters involve a number of spins
growing as $L^2$, we can identify $\xi_B(B_J^*)$
with $L$. Let $B^*$ be the disorder average of $B_J^*$; then
$B^* \sim L^{-y_B}$ from which we can estimate $y_B$. We find that a pure
power with $y_B$ set to its value in the droplet picture describes
the data quite well; in the inset of Fig.~\ref{fig:fig4}
we display the product $L^{1.282} B^*$ 
as a function of $1/L$ and see that the behavior is compatible
with a large $L$ limit with $O(1/L)$ finite size
effects. Direct fits to the form $B^*(L) = u L^{-y_B} (1 + v/L)$
give $y_B$s in the range $1.28$ to $1.30$ depending on the 
points included in the fit.

Given the magnetic length, one can perform finite size scaling (FSS) on the
magnetization data $m(B,L)$. Since FSS applies to the singular
part of an observable, we should have a data collapse according to
\begin{equation}
\frac{m(B,L)-\chi_1 B}{m(B^*,L)-\chi_1 B^*} = W(B/B^*)
\end{equation}
$W$ being a universal function, $W(0)=O(1)$ and
$W(x)\sim x^{1/\delta}$ at large $x$.
Using the value of $\chi_1$ previously determined,
we display in Fig.~\ref{fig:fig4} 
the associated data. The collapse is excellent and we have checked
that this also holds when the $J_{ij}$ are drawn
from an exponential distribution. Added to the figure is the
function $x^{1/\delta}$ to guide the eye ($\delta=1.282$ as 
predicted by the droplet model).

\paragraph*{Conclusions ---}
We have investigated the $2d$ Ising spin glass with Gaussian and
exponential couplings
at zero temperature as a function of the magnetic field.
The magnetization exponent $\delta$ can be measured; previous
studies did not find good agreement with the droplet model
prediction $\delta=1.282$ because the analytic contributions
to the magnetization curve were mishandled, while in this work we
found instead $1.28 \le \delta \le 1.32$.
We also performed a direct measurement of the magnetic length,
obtaining for the associated exponent $1.28 \le y_B \le 1.30$, again in
excellent agreement with the droplet prediction.
With this length we showed that finite size scaling is 
realized without going to infinitesimal fields or
huge lattices.
Finally, we validated the hypotheses underlying
the arguments of the droplet model inherent to the
in-field case; we find in particular that in the low-field 
limit the spin clusters
that are relevant are compact and have random magnetizations.
In summary, by combining improved computational techniques
and greater care in the analysis, we have lifted
the discrepancy on the magnetic exponents
that has existed for over a decade
between numerics and droplet scaling.

We thank T. Jorg for helpful comments.
The computations were performed on the cliot cluster of the
Regional Computing Center and on the scale cluster of E.
Speckenmeyer's group, both in Cologne.
FL has been supported by the German Science Foundation in the projects
Ju 204/9 and Li 1675/1 
and by the Marie Curie RTN ADONET 504438 funded by the EU.
This work was supported also by the EEC's 
HPP under contract HPRN-CT-2002-00307 (DYGLAGEMEM).

\bibliographystyle{apsrev}

\bibliography{/home/martino/Papers/Bib/references,/home/martino/Papers/Bib/co}

\end{document}